\documentclass[%
reprint,
superscriptaddress,
showpacs,
nofootinbib,
nobibnotes,
 amsmath,amssymb,
 aps,
prl,
floatfix,
]{revtex4-1}

\usepackage{graphicx}
\usepackage{dcolumn}
\newcolumntype{.}{D{.}{.}{-1}}
\usepackage{bm}
\usepackage{float,array,booktabs,amsmath,amssymb}

\begin{document}

\preprint{APS/123-QED}
 \title{A random walk Monte Carlo simulation study of COVID-19-like infection spread
}  

%
\author{S.~Triambak}
\email{striambak@uwc.ac.za}
\affiliation{Department of Physics and Astronomy, University of the Western Cape, P/B X17, Bellville 7535, South Africa}%
\author{D.\,P.~Mahapatra}
\email{dpm.iopb@gmail.com}
\affiliation{Department of Physics, Utkal University, Vani Vihar, Bhubaneshwar 751004, India}
%
%
%
 
\date{\today}

\begin{abstract}
Recent analysis of early COVID-19 data from China showed that the number of confirmed cases followed a subexponential power-law increase, with a growth exponent of around 2.2 [B.\,F.~Maier, D.~Brockmann, {\it Science} {\bf 368}, 742 (2020)]. The power-law behavior was attributed to a combination of effective containment and mitigation measures employed as well as behavioral changes by the population. In this work, we report a random walk Monte Carlo simulation study of proximity-based infection spread. Control interventions such as lockdown measures and mobility restrictions are incorporated in the simulations through a single parameter, the size of each step in the random walk process. The step size $l$ is taken to be a multiple of $\langle r \rangle$, which is the average separation between individuals. Three temporal growth regimes (quadratic, intermediate power-law and exponential) are shown to emerge naturally from our simulations. For $l = \langle r \rangle$, we get intermediate power-law growth exponents that are in general agreement with available data from China. On the other hand, we obtain a quadratic growth for smaller step sizes $l \lesssim \langle r \rangle/2 $,  while for large $l$ the growth is found to be exponential. 
We further performed a comparative case study of early fatality data (under varying levels of lockdown conditions) from three other countries, India, Brazil and South Africa. We show that reasonable agreement with these data can be obtained by incorporating small-world-like connections in our simulations. 
\end{abstract}
\maketitle
%
\paragraph*{Introduction} Following its outbreak in the Hubei province of China, the global spread of the novel coronavirus disease (COVID-19) has reignited efforts to better understand infection spread and mortality rates during the pandemic. Significant emphasis was placed on modeling the spatio-temporal spread of the disease, in order to make reliable predictions. A key statistic in such epidemiological analysis is the basic reproduction number $R_0$, which defines the expected number of secondary cases from one infected individual 
in a completely susceptible population. Data from the very initial phase of the COVID-19 outbreak showed good agreement with models that assumed an exponential growth of infections in time $(t)$, with a mean $R_0$ ranging from $2.24$ to $3.58$~\cite{zhao,zhou}. However, subsequent laboratory confirmed cases in Hubei showed that soon after the initial stage, the temporal growth in the cumulative number of infections ($N$) was instead subexponential and agreed reasonably well with a power-law scaling $N \propto t^\alpha$~\cite{maier}. This was consistent with data from other affected regions in mainland China (with $\alpha =  2.1\pm 0.3$) and was attributed to a depletion of the susceptible population due to effective containment and mitigation strategies that were put in place and followed after the initial unhindered outbreak~\cite{maier}.

A potential stumbling block in such analyses is that the reported number of infected cases may be inaccurate, due to a non-uniform sampling of the entire susceptible population in a given region. In such a scenario one can alternatively examine the number of reported deaths (due to COVID-19 complications) as a function of time. This is justified, as the number of deaths are generally more accurately recorded and (under non-variable containment, mitigation and treatment strategies) can be assumed to be a fixed fraction of the total infected population. 
Indeed, early mortality data from the National Health Commission of the People's Republic of China and Health Commission of the Hubei Province showed similar power-law behavior, with an exponent $\alpha \approx 2.2$~\cite{mingli} that agreed with the observations in Ref.~\cite{maier}. Independently, it has been proposed that the near-quadratic power-law scaling of the cumulative number of deaths (infections) in China can be explained with an epidemiological model that allows `peripheral spreading'~\cite{axel}. In this model, once infections are identified in a location (labeled as a `hotspot'), and the subpopulation from the region is isolated, the growth of infections within this confined local community rises exponentially until no further infections are possible. Once this saturation is reached, further spread of the disease to outside the region is inevitable, due to interactions at the periphery of the confined population. The growth of infections due to such peripheral spreading is shown to be quadratic in time and agreed piecewise with the data from China~\cite{axel}. 

In light of the above, we performed a random-walk Monte Carlo simulation study of the spread of a highly infectious disease such as COVID-19, with particular emphasis on its temporal growth within a constrained population. We show  commonalities between independent models describing such COVID-19 growth, while simultaneously demonstrating the efficacy of the random-walk model to make predictions. This work also complements other studies of infectious disease spread through transmission networks, such as with aviation~\cite{Hufnagel}, currency dispersal~\cite{Brockmann:06} and mobile phone~\cite{Schlosser} data. Our model, described below, has the ability to capture random interactions that may be missed in such data-driven contact network studies and is relatively easy to access compared to most models that study the spread of epidemics. Additionally, we show that our simulations have the ability to compartmentalize the data, similar to susceptible-infectious-recovered (SIR) or susceptible-infectious-recovered-susceptible (SIRS) type models~\cite{Bailey_book}, that are conventionally used in the study of infectious disease spread.

\paragraph*{Monte Carlo simulations}
Random walks, particularly on a lattice, have been extensively studied in the past~\cite{mccrea,montroll,masolivier,benjamini,batchelor}. Similar studies have also been used to analyze contact interactions~\cite{Harris} as stochastic processes, to better understand epidemic spread~\cite{Mollison,Filipe1,Filipe2,Liggett,Draief,Bestehorn,Kiss,Allen}. In such analysis, one often has to rely on certain approximations~\cite{Filipe1,Filipe2,Filipe3,Levin,Ellner}, due the complexity in describing stochastically interacting populations over a geographical region. In this regard, Monte Carlo simulations offer a viable alternative and are pursued using different approaches. For example, early work~\cite{Bailey} assumed each susceptible individual to occupy a point on a square lattice, having a certain probability to contract a disease from an infectious neighbor, who may be located in one of its nearest lattice positions. In other simulations, the population was distributed in a grid of cells~\cite{Bartlett,Kelker}, and similar to percolation models, the stochastic movement of infectives as well as susceptibles between cells with common boundaries resulted in the spread of the disease. More recently, a dynamical network random-walk model~\cite{Frasca,Buscarino} was used to study the effects of long-ranged spatial mobility on epidemic spread. Our simulations are along similar lines. Here, the individuals of an entirely susceptible population are described as identical and independent random walkers, represented by uniformly distributed random points in an isolated two-dimensional region. The simulations begin with an initial condition of one infected walker near the center, assuming all other points are `normal' (uninfected). As the simulation progresses, all individual walkers take simultaneous steps in random directions. For cases when walkers step outside the bounded area, a boundary condition was imposed so that the transgressing coordinates were reflected back into the bounded region. 
The incremental number of simultaneous discrete steps taken by the random walkers quantifies both time progression as well as spatial mobility.  Similar to Ref.~\cite{larralde}, we call these increments `time-steps'. 
%
\begin{figure*}[t]
  \centering
    \includegraphics[width=.95\textwidth]{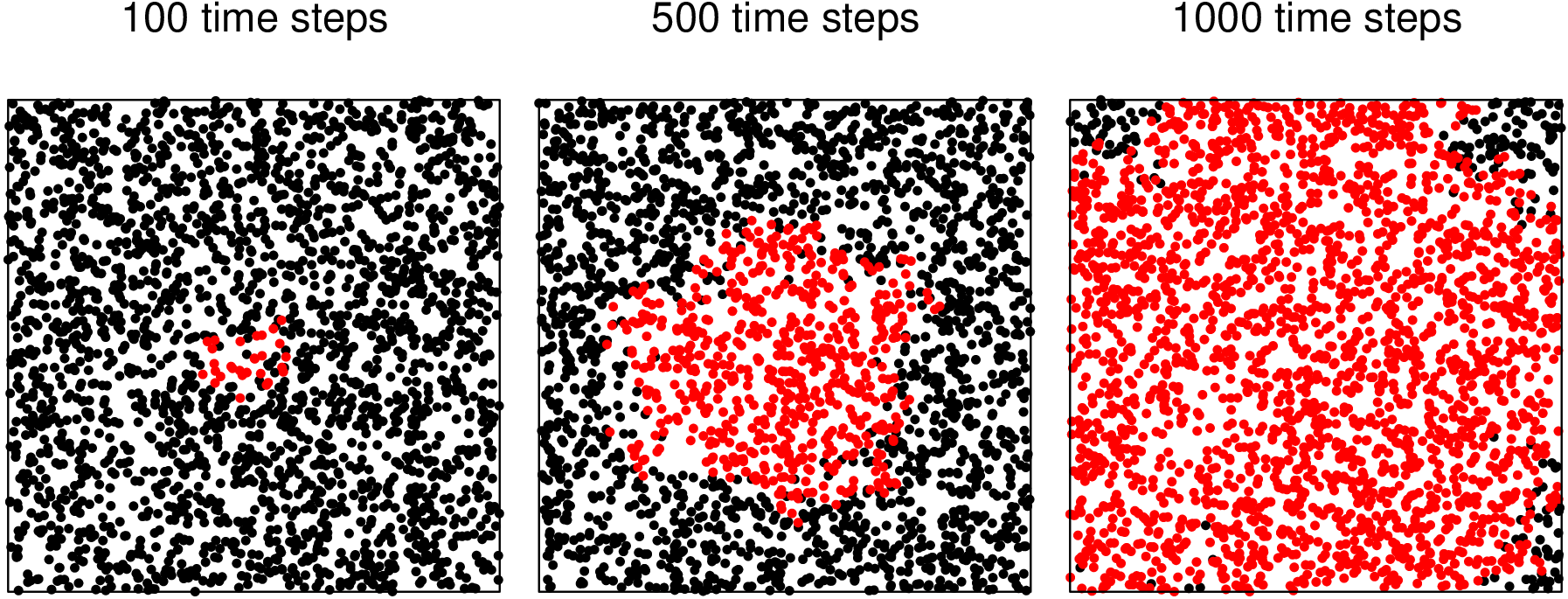}
\caption{
An example of proximity-based infection spread obtained using the random walk Monte Carlo simulations described in this work. Each of the panels shown above has a population of 2.5k over a unit area. The average distance $\langle r \rangle$ between any two points is  = 0.02 units.  In this case every point (walker) takes randomly directed steps of length $l = 0.25\langle r \rangle$. Further details are described in the text below.  
}
  \label{fig:dotplot}
    \end{figure*}
%
The `spread' of the disease is said to occur whenever an infected point comes within a `touching' distance from a `normal' susceptible point, thereby passing on the disease. 
In such a manner, the growth in the number of infected points with respect to the number of discrete time-steps ($t$) determines the temporal evolution of the infection spread.
%

To put the above in perspective, for $N$ randomly distributed points over area $A$, the mean distance of separation $\langle r \rangle$ between any two random walkers is  $\sim \sqrt{A/N}$. Therefore, for a metropolis such as New York city, which has a population density of $\sim$~10,000/sq. km, $\langle r \rangle$ is $\sim$10~m. For an arbitrarily sized region, given the fixed population density, this would result in 10,000 points per unit area, with $\langle r \rangle = 10^{-2}$ length units. We assume a `touching' separation of~2~meters, which is the nominal safe distance recommended in most countries. This corresponds to $2\times10^{-3}$ normalized length units. If the distance between any infected data point and a normal (susceptible) point is less than or equal to this value, the normal point is flagged as infected in the simulations. To illustrate the above, we show an example of such spatio-temporal disease progression in Fig.~\ref{fig:dotplot}, where the flow of time is quantified in terms of the number of `time-steps' in the simulation.   
We discuss specific results from three sets of simulations below.  The first two assume a synchronous SI (susceptible-infected) model, in which all `normal' points are 100\% susceptible, while the third (synchronous SIRS) set of simulations assumed small recovered fractions of the population, some of whom are susceptible to reinfection.        

\paragraph{Simulation set I (Fixed population density, fixed step length, different population sizes)} These simulations investigated the spread of infection in the hypothetical metropolis mentioned above\footnote{All simulations described here were carried out for $N$ random walkers over a unit area.}, assuming that each walker's spatial mobility is effectively constrained due to containment (lockdown) measures. It is intuitively reasonable to assume that such a restriction can be achieved by imposing a condition that all members of the populace only take random steps of length $l = \langle r \rangle$, where $\langle r \rangle$ is the average distance between individuals. This   
ensures  
each random walker to be confined (on average) within a local neighborhood. We performed three such simulations for a fixed density of 10k/unit area and three population sizes 10k, 6k and 2.5k respectively. 

\begin{figure}[t]
\centering
    \includegraphics[scale = 0.37]{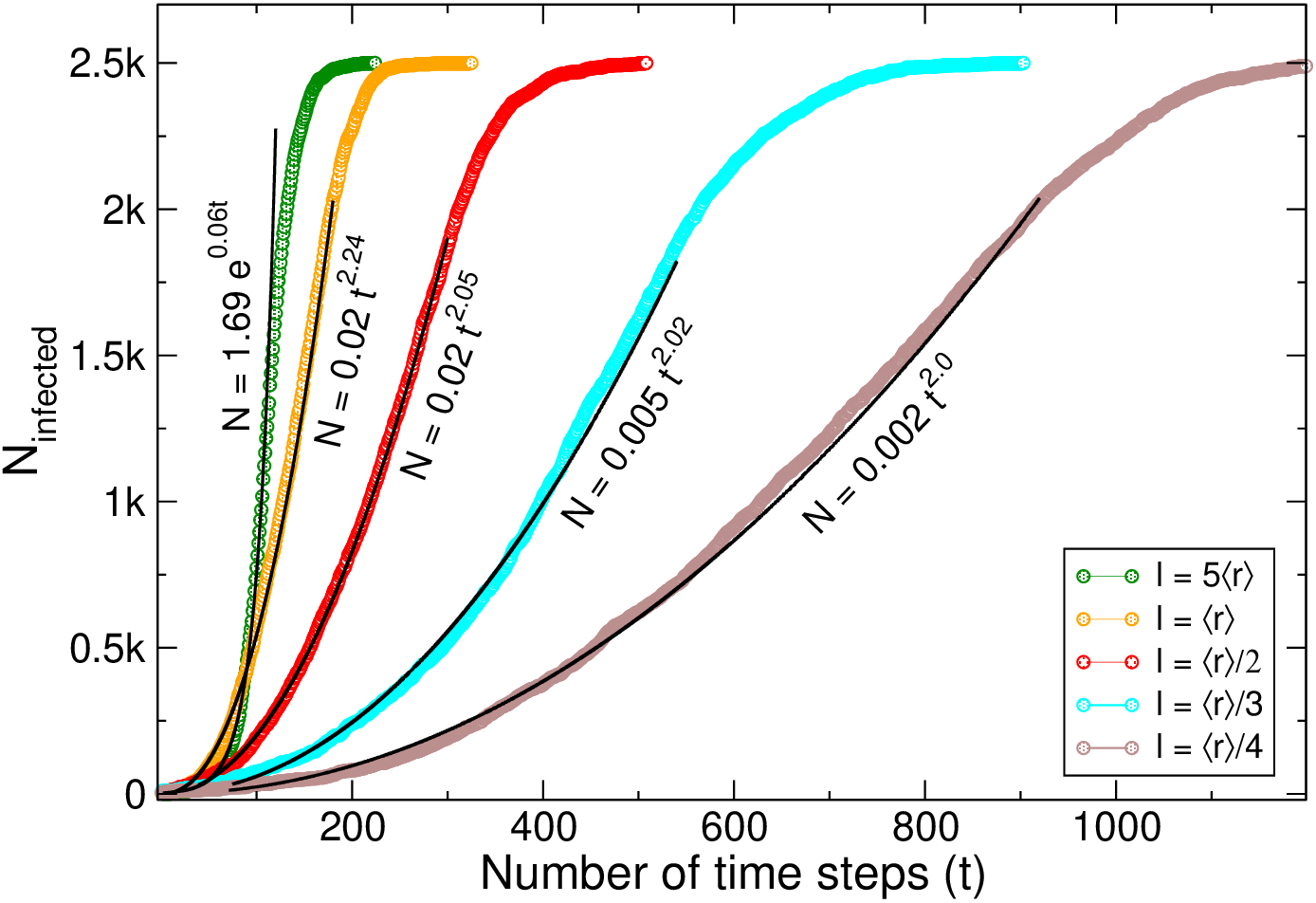}
\caption{Simulated growth in the number of infections and their best-fit curves, obtained for different step sizes 
taken by the random walkers in a population of 2.5k/unit area. Below a threshold value of $ l \sim \langle r \rangle$/2, the growth is observed to be \textit{nearly} quadratic, regardless of the step size. The other extreme shows exponential behavior, while power-law growth similar to what was observed in China~\cite{maier} lies in the intermediate regime. Nearly identical trends are observed for the 10k case.} 
    \label{fig:regimes}
    \end{figure}

\paragraph{Simulation set II (Fixed population and density, different step lengths)} In the next step we probed the dependence on both population density and $l$ in five separate subsets of simulations. These simulations assumed densities of 10k and 2.5k walkers/unit area, and different step sizes for the walkers, with lengths  $\langle r \rangle/4$, $\langle r \rangle/3$, $\langle r \rangle/2$, $\langle r \rangle$ and $5\langle r \rangle$. The 2.5k results are shown in Fig.~\ref{fig:regimes}. Further discussion follows in Section~\ref{sec:results}. 

\paragraph{Simulation set III (Fixed population, density and step lengths, recovery and reinfection allowed)} In these SIR and SIRS variants, we studied the effects of a small recovery and reinfection rate within a fraction of the population and their effect on the growth exponent. We independently investigated scenarios with recovery percentages of 0.02\%, 0.1\% and 0.5\% (similar to Ref.~\cite{axel}), such that i) all the recovered individuals are immune and ii) a randomly selected 5\% of the recovered population are susceptible to reinfection.  
\begin{figure*}[t]
  \centering
    \includegraphics[width=.95\textwidth]{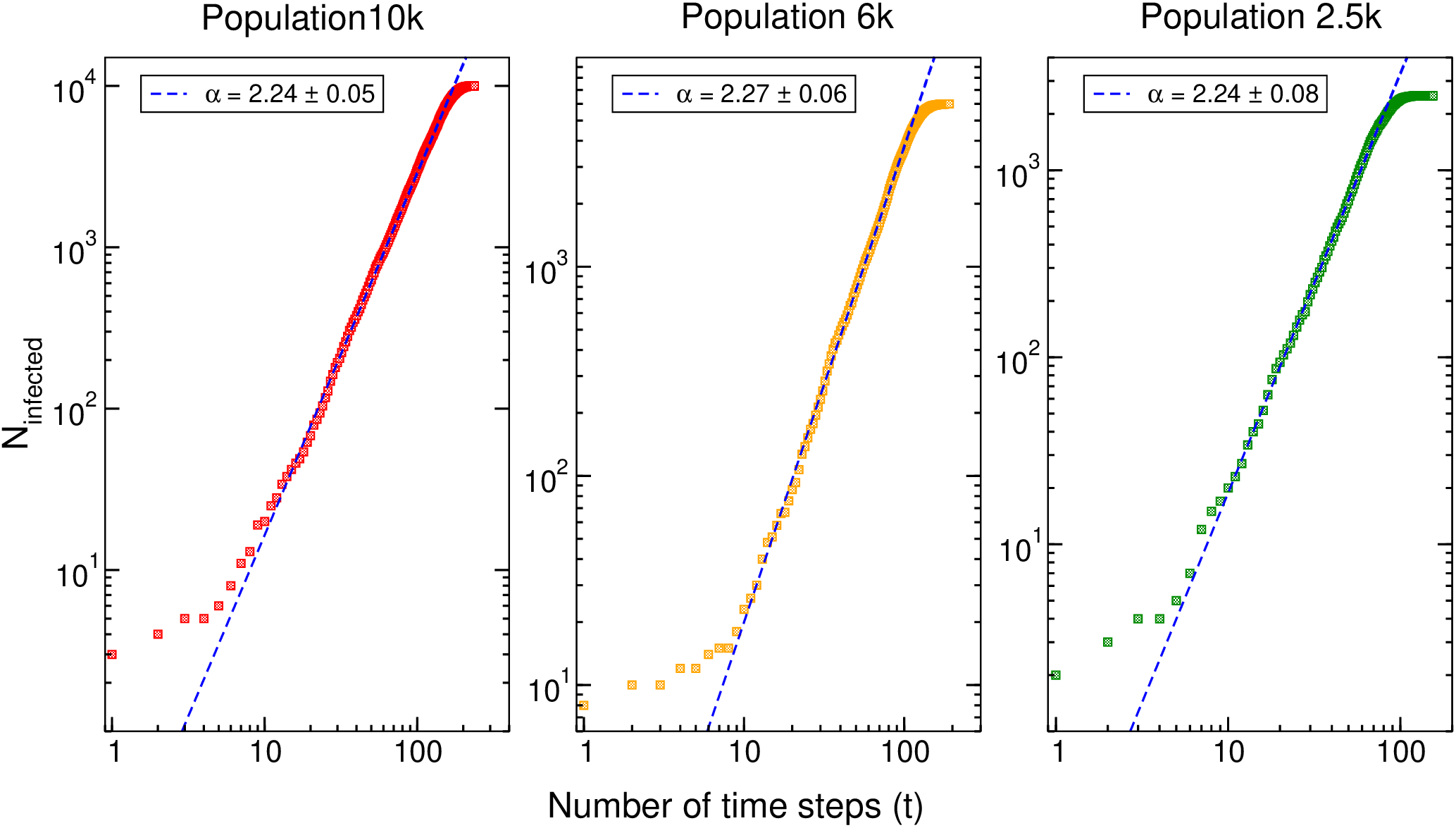}
\caption{Simulation results for three population sizes with a fixed number density of 10k/unit area. The simulations assume that all random walkers take steps of length $\langle r \rangle$, their mean separation distance. The dashed lines are best-fit results for power-law growth $N = At^\alpha$. The quoted uncertainty for each extracted $\alpha$ includes a $\pm 1\sigma$ statistical uncertainty and a systematic contribution. The latter were estimated by applying a conservative one channel shift to the data along the time axis and redoing the fits. 
}
  \label{fig:sims}
    \end{figure*}

\begin{figure}[t]
  \centering
    \includegraphics[scale = 0.21]{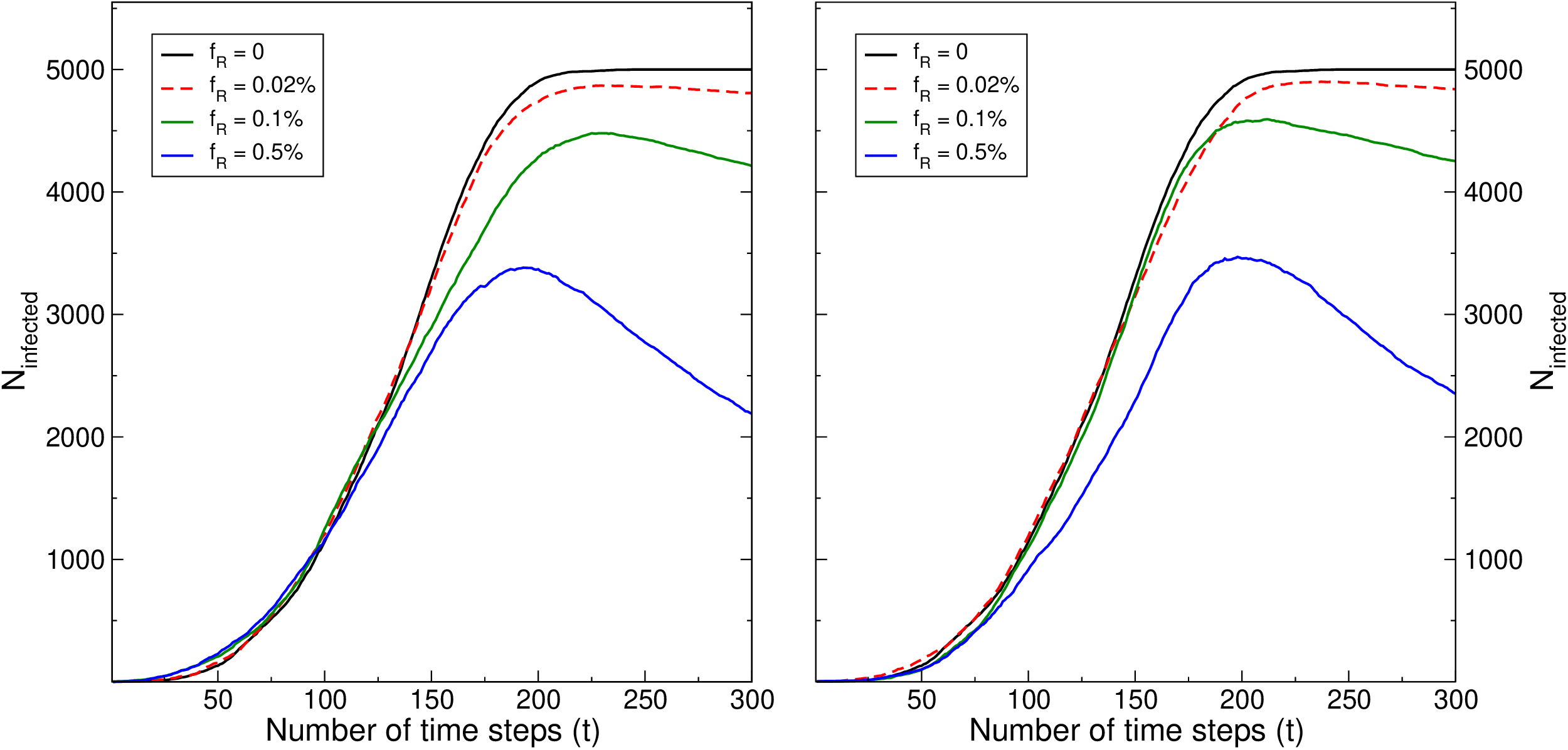}
\caption{Left panel: SIR-type simulation results obtained for different values of $f_R$, the recovered fraction. Right panel: SIRS-type simulation results obtained assuming a random $5\%$ of $f_R$ are susceptible to reinfection.
}
  \label{fig:brand}
    \end{figure}

\paragraph*{Results and analysis}
\label{sec:results}
In Fig.~\ref{fig:sims} we plot the growth in the cumulative number of infected points, obtained from simulation set I. The results show that independent of population size, the number of infections follow a $t^\alpha$ power-law growth in time, with $\alpha$ about 2.2.
While the power-law behavior may not be completely unexpected~\cite{meyer}, it is interesting that we obtain very similar values of near-quadratic exponents, as observed with the data reported in Refs.~\cite{maier,mingli}. In simulation set II, for step length $\langle r \rangle$, we determine almost identical power-law growth as in Fig.~\ref{fig:sims}, again in agreement with the observations of Refs.~\cite{maier,mingli}. This is shown\footnote{We do not quote uncertainties in the fit parameters here, as this figure only serves to highlight the \textit{systematic trends} of the curves.} in Fig.~\ref{fig:regimes}. Our extracted power-law exponents are consistently similar for this step size, regardless of the population density used.  In comparison, if all members of the sample population were to take larger random steps of length $5\langle r \rangle$, on average interacting with points located  further away than their nearest neighbors, we find that the number of infected individuals blows up rapidly, showing near exponential behavior. This would be similar to a scenario where no control interventions are in place or being followed. Not surprisingly, the slope for exponential growth is found to strongly depend on the population density, and is larger at higher densities.
Figure~\ref{fig:regimes} also shows the other extreme in terms of the temporal growth, obtained using step sizes smaller than $\langle r \rangle$. 
As apparent in the figure, the results from these simulations show near-quadratic growth\footnote{One would notice that the quadratic power-law fits show better agreement with the data generated using longer step-lengths, compared to the ones corresponding to $l = \langle r \rangle/3$ and $\langle r \rangle/4$.  We note that this is mainly due to statistics, on account of the small step-size used in the simulations. We further re-emphasize that Fig.~\ref{fig:regimes} only serves to highlight the \textit{trends in the growth curves}, which are {\it nearly} quadratic for $l \lesssim \langle r \rangle/2$.}, for all step sizes less than a threshold value of around $\langle r \rangle$/2.  This effectively implies a lower-bound on the growth exponent ($\alpha = 2$), exactly as in the case of peripheral spreading~\cite{axel}. Furthermore, the effect of mobility restriction is clearly evident from the observed delay in reaching the saturation value, when smaller step-lengths are used. Finally, our results for simulation set III show that a SIR recovery fraction of the order $0.02\%$ does not affect the growth exponent significantly. We find that a small reinfection component in our SIRS-type simulations leads to reasonable agreement with temporal growth from the SI results, even when the recovered fraction is comparatively larger, at around $0.1\%$. This is shown in Fig.~\ref{fig:brand}, whose results were obtained for 5k random walkers/unit area.
In light of the above, we assume that (small) recovery and reinfection rates do not play a significant role in our interpretation of results.      
%

The three growth regimes (quadratic, intermediate power-law and exponential), obtained by changing the random walkers' mobility through their step sizes are closely linked to other models. We observe that the slowest unavoidable temporal growth is quadratic in nature. This is not surprising for an unbiased and uncorrelated random walk on a plane. For each random walker, the root-mean-square (rms) displacement after taking $t$ steps of fixed length $l$ goes as $l\sqrt{t}$. Therefore the probability of a susceptible individual intercepting a single infectious walker is proportional to the overlapping area covered by {\it both} of them, which scales as $t^2$. This is similar to the peripheral spreading model proposed in Ref.~\cite{axel}, where the disease spread to the outside of an isolated population is through $n$ infectives located in a narrow band at the circumference of the confined region. In such a scenario $n$ scales as $\sqrt{N}$, where $N$ is the total number of infected people at that time. This results in quadratic growth, with $N(t) \propto t^2$~\cite{axel}. At the other extreme, for long step lengths taken by the walkers in our simulations, the growth is found to be exponential, and consistent with what one would expect from a homogeneous mixing~\cite{Fofana} of the population. Such exponential growth is predicted by compartmental models~\cite{Bailey_book}, that also allow a `diffusion' of the disease~\cite{Noble,Kallen} through random walks of the population, provided that there is no depletion of the susceptible population~\cite{Bailey_book} and that intervention measures/behavioral changes are not enforced or followed. Finally a step-length parameter of size $\langle r \rangle$ for the random walkers reproduces the intermediate power-law growth curves from China reported in Refs.~\cite{maier,mingli}. This is consistent with Ref.~\cite{maier} that used a modified compartmental model, which took into consideration both quarantine procedures as well as containment strategies. It is interesting that a simple logarithmic correction to quadratic growth (so that $t^2 \to t^2 \ln t$) results in a power-law exponent of about 2.5, that is in rough agreement with the intermediate values for contained growth reported from China~\cite{maier,mingli}.

We further note that since quadratic scaling appears to be the limiting case, it is unrealistic for a large population to achieve such minimal growth. Therefore, at face value the available data suggest that the containment measures and response in the eight affected Chinese provinces mentioned in Ref.~\cite{maier} most likely could not have been significantly improved upon. Furthermore, for both quadratic and intermediate power-law growth, the growth exponent is found to be independent of the population density in our simulations. This is not unexpected when the step size $l$ scales as $\langle r \rangle = \sigma^{-1/2}$, where $\sigma$ is the population density. Any change in $\sigma$ would be offset by a corresponding change in $l$ for the random walkers.    

\paragraph*{Power-law, exponential growth and small-world-like connections in observed data (India, Brazil and South Africa)}
\begin{figure*}[t]
  \centering
    \includegraphics[width=.99\textwidth]{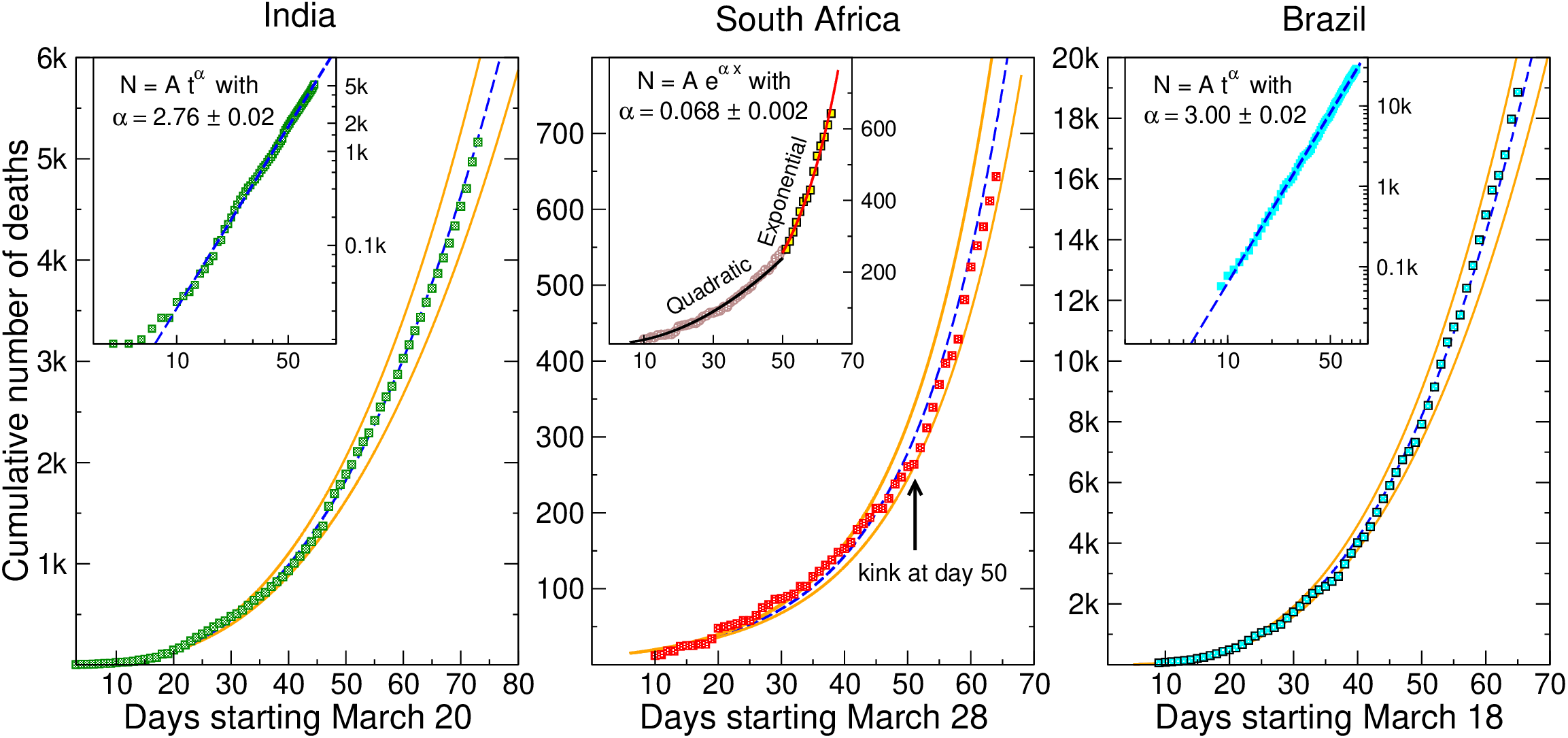}
\caption{Cumulative number of deaths reported for India, South Africa and Brazil from the WHO, until June~1~\cite{who}. The fitted growth curves are shown together with their 95\% C.L. bands. The Brazil and India data show a single growth-exponent, consistent with power-law behavior. On the other hand the overall data for South Africa suggests exponential growth. Closer inspection shows that the initial growth phase for South Africa was quadratic. The growth exponents are quoted in the insets with $\pm 1\sigma$ uncertainty.}
  \label{fig:who}
    \end{figure*}
To make further comparisons, we examine the daily growth in the cumulative number of deaths reported~\cite{who} for three countries, South Africa, India and Brazil, until June 1, 2020. These BRICS countries were not arbitrarily chosen, as they present an interesting comparative study for several reasons. They face common challenges in terms of poverty and economic inequality, and are home to some of the most densely populated informal settlements in the world (such as Khayelitsha in Cape Town, Dharavi in Mumbai and the favelas of Rio de Janeiro and S\~ao Paulo). The lack of proper sanitation is a common theme in these impoverished city pockets, where, given the circumstances, expecting the residents to follow strict social-distancing protocols is a tall order~\cite{godfrey,slum,brazil}. Secondly, the response of the political leadership of Brazil to the COVID-19 crisis was strikingly different from the governments of India and South Africa. While the latter two countries swiftly imposed extended periods of severe lockdown~\cite{lk1,lk2,lk3,lk4} starting in the month of March (2020), Brazil did not pursue a concerted policy for such containment~\cite{sowhat}. 
The cumulative death data reported for the three countries, with their corresponding fits are plotted in Fig.~\ref{fig:who}. While we do observe power-law growths with exponents of $2.76 \pm 0.02$ and $3.00 \pm 0.02$ for India and Brazil, the growth curve for South Africa is surprisingly much steeper. As shown in its inset, the growth was nearly quadratic for a significant portion of the time, following which there is a steep exponential rise starting around day 50 from March 28. 
It is worthwhile to note at this point that the most stringent lockdown measures (at Level~5) were imposed in South Africa until May~1~\cite{lk4}. The restrictions were only slightly relaxed after that, to Level~4 during the month of May. Interestingly, the data show that the exponential growth begins around May~17. Given that the coronavirus has an approximately two week incubation period, the above observation suggests that the two largely different growth exponents for South Africa are most likely due to a modest containment of the disease under Level~4 lockdown. 
\begin{figure}[t]
\centering
    \includegraphics[width=.46\textwidth]{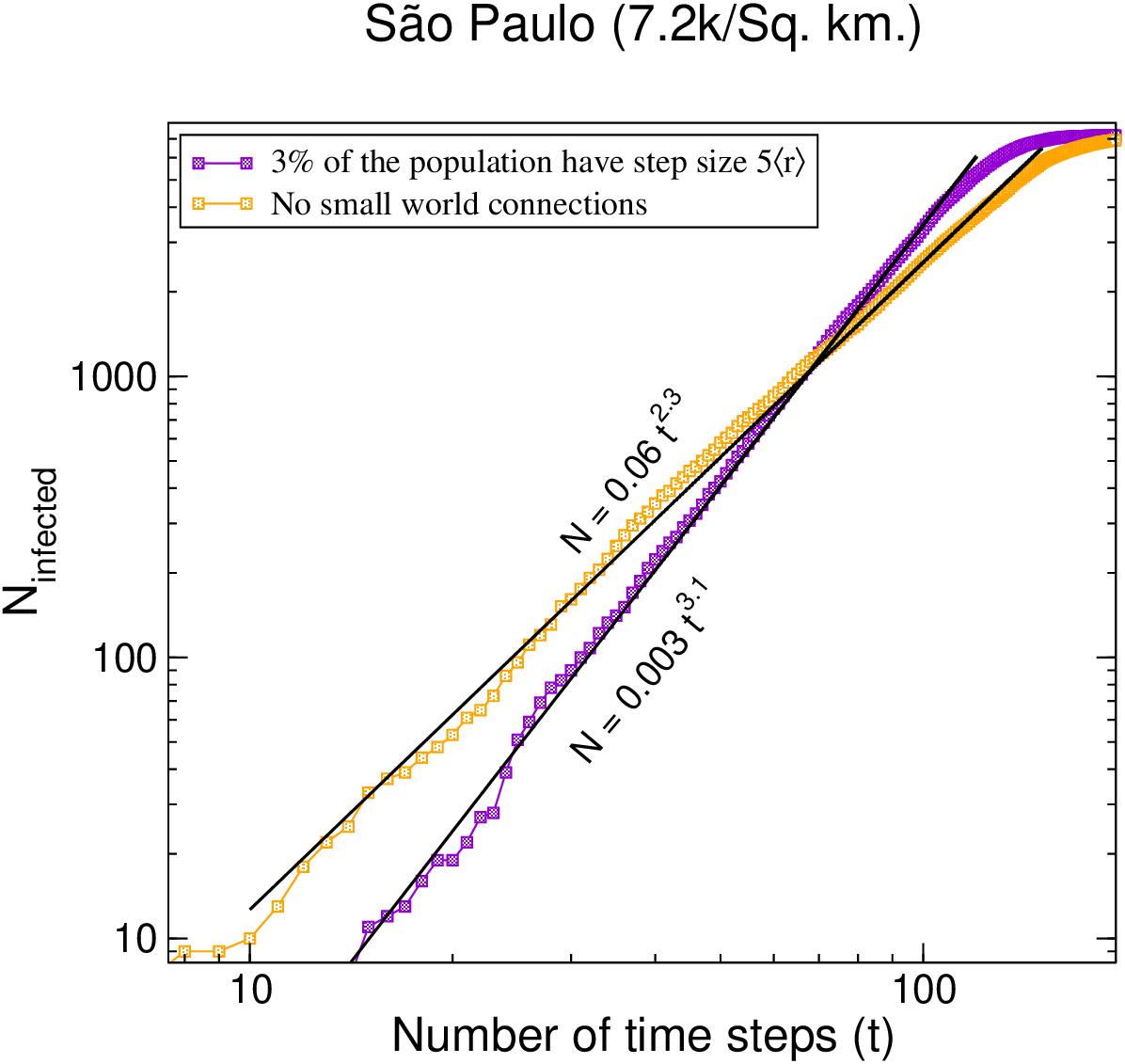}
\caption{Comparison of power-law growth exponents for an example city such as S\~ao Paulo (population density 7.2k/unit area), with and without small-world-like connections in the simulation. The latter were introduced by having a random 3\% of the population take steps of length $5\langle r \rangle$. The uncertainties in $\alpha$ are quoted similarly as in Fig.~\ref{fig:sims}. } 
    \label{fig:sao}
    \end{figure}

While the observed growth exponent for the number of fatalities in Brazil is not unexpected, how does one explain the higher-than-anticipated growth exponents for South Africa and India? We show below that an extension to our model, along the lines of a small world network~\cite{watts,ziff} can explain the observed growth. For this, we took into consideration a more realistic lockdown scenario that includes a small number of outliers in the population (representing essential service workers and non-compliant citizens etc.), who are allowed to take much larger randomized steps, bounded by the sampling area $A$. 

If strict containment measures were not adhered by members of the population, it would correspond to a combination of two effects in our random walk model: (i) All the random walkers use relatively larger step sizes. (ii) A small fraction of the population has much longer-ranged mobility compared to the above. This establishes small world connections between infected individuals and the rest of the susceptible population. 
%
%
    
Our simulation results for an example city such as S\~ao Paulo (with a population density of 7,200/unit area) are shown in Fig.~\ref{fig:sao}. For a uniform step length $l = \langle r \rangle$ we determine a power-law exponent of $2.30 \pm 0.06$, again in agreement with our previous observations. On increasing the step lengths of a randomized ensemble comprising 3\% of the city's population to $l = 5\langle r \rangle$, we find that the exponent increases to $3.09 \pm 0.08$, very similar to the Brazil data shown in Fig.~\ref{fig:who}. The higher growth exponent for the data from India can be explained similarly. 
Despite its best attempts, the country's COVID-19 containment strategy was challenged by the sheer scale and diversity of its population. For example, it is known that on several occasions people defied social-distancing measures to attend religious gatherings in large numbers~\cite{india1,india2}. Furthermore, the sudden and unprecedented lockdown in India resulted in a humanitarian crisis, with millions of daily wage inter-state migrant workers from the rural hinterland left jobless in the big cities~\cite{mig1}. This led to a large-scale migration back home for thousands of such families, many of them traversing large distances of the country on foot~\cite{mig2,mig3}. 
The above clearly shows the contribution of long distance movers to the spread of the pandemic.  It is well known that long-ranged dispersal can dramatically accelerate the spread of infection~\cite{fisher}.  

%
%
\begin{figure}[t]
\centering
    \includegraphics[scale = 0.34]{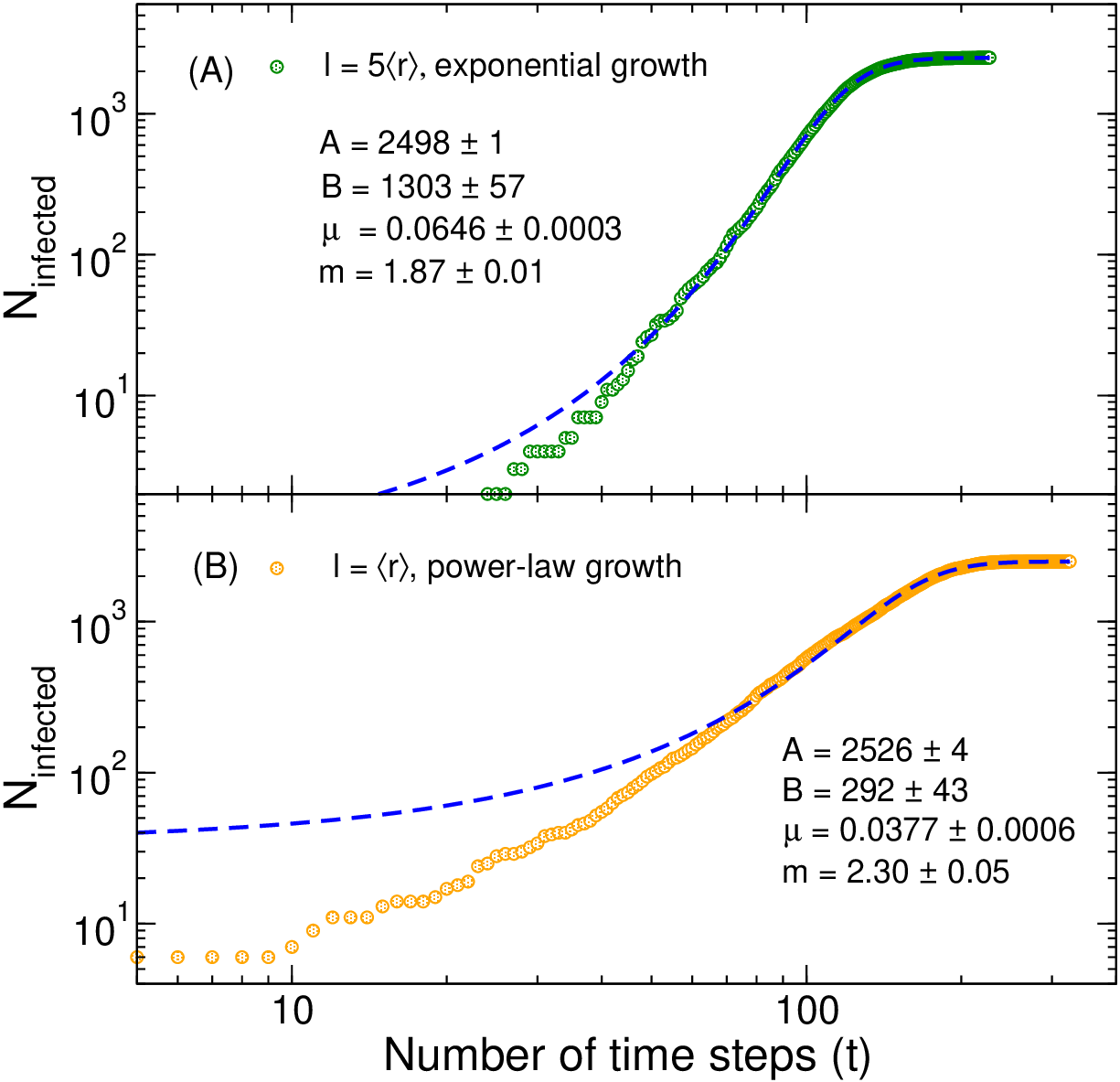}
\caption{Monte Carlo results for (A) exponential and (B) constrained growth (both with saturation) fitted to a Richards logistic function $N(t) = A(1+Be^{-\mu t})^{1/(1-m)}$. The  data are the same as shown in Fig.~\ref{fig:regimes}
}
\label{fig:fits}
    \end{figure}%

\begin{figure}[b]
\centering
    \includegraphics[scale = 0.34]{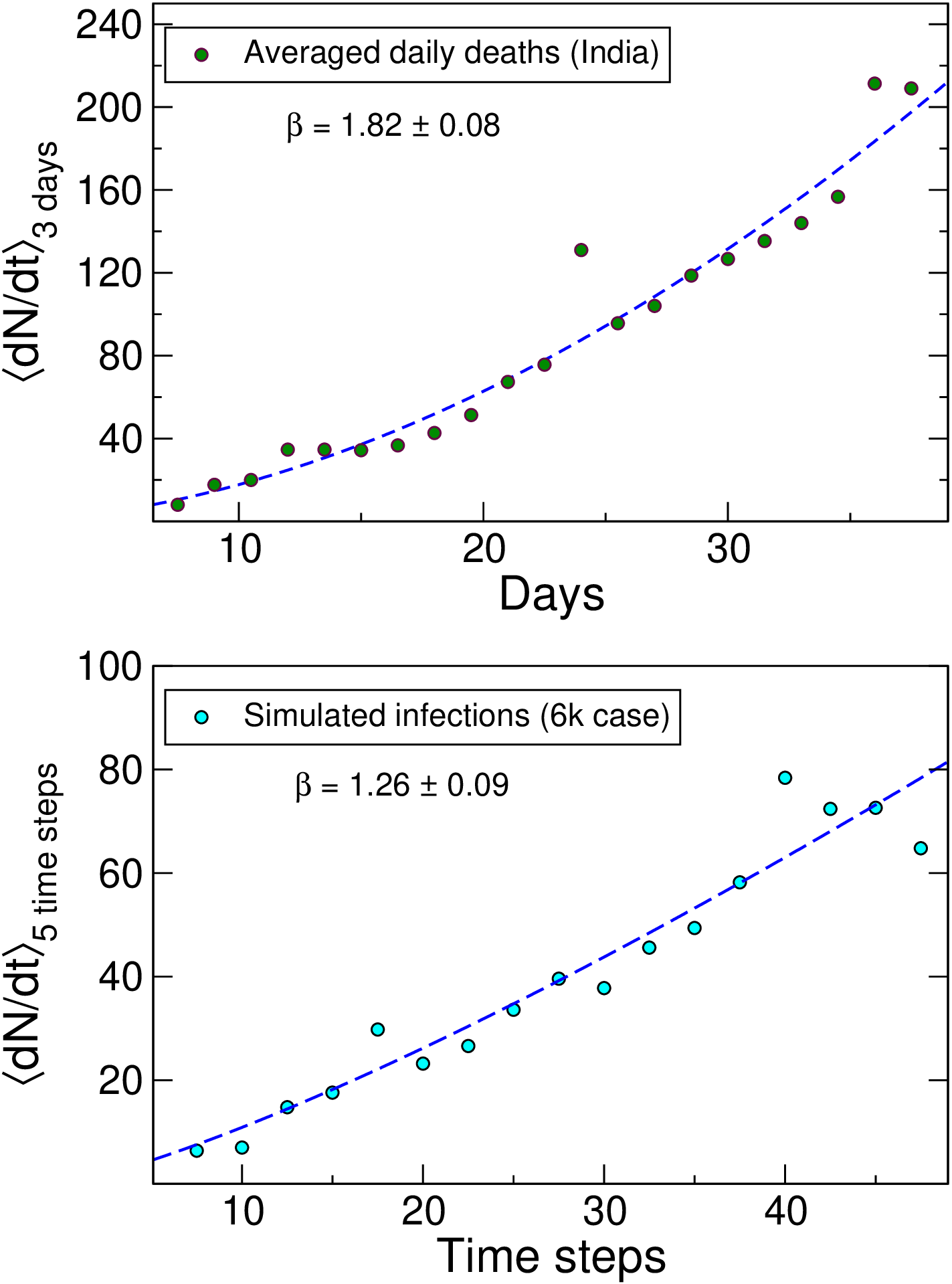}
\caption{Top panel: Growth in three-day-averaged daily deaths reported from India and its corresponding power law fit $\langle dN/dt \rangle = At^\beta$. Bottom panel: Similar data obtained from our random walk Monte Carlo simulations, for a city with population of 6k and density 10k/unit area. In both plots 
the $\beta$ are quoted with $\pm 1\sigma$ uncertainty.} 
\label{fig:daily}
    \end{figure}

Recently, there have been several attempts~(see Ref.~\cite{singer} and references therein) to fit the sigmoid-type curves for country specific COVID-19 infections with logistic growth models, including a generalized logistic function of type \mbox{$N(t) = A(1+Be^{-\mu t})^{1/(1-m)}$}, that solves the Richards differential equation~\cite{richards}. We caution that such an approach can lead to inaccuracies, particularly when an effective containment policy is followed. 
It is clear that the above expression for $N(t)$ does {\it not} produce a linear relationship between $\ln N$ and $\ln t$, as expected for (contained) power-law growth.
This is manifested in the results of our simulations and validates our Monte Carlo random walk approach.  
To illustrate the above, we show in Fig.~\ref{fig:fits} simulation results for random walkers with both unconstrained and constrained mobility, generated with step lengths $l = 5\langle r \rangle$ and $\langle r \rangle$ respectively. While the generalized logistic function provides a reasonable fit for the unconstrained curve (exponential growth), a large discrepancy is observed in the other (power-law) case, with significantly different values for the fit parameters. 
For $t^\alpha$ type power-law growth, it is apparent that the infection rate $dN/dt$ should be proportional to $\alpha t^{\alpha-1}$.
This is supported by our simulated data. As an example, we show data corresponding to three-day averaged values for the reported daily deaths from India and their corresponding power-law fit in the top panel of Fig.~\ref{fig:daily}. As expected, we obtain a growth exponent of $\beta = \alpha-1$ (for $\alpha = 2.8$). The bottom panel in the same figure shows a similar analysis performed for our simulated data, obtained for a population of 6k and density of 10k/unit area. These data (which are the same as presented in the central panel of Fig.~\ref{fig:sims}) show exactly the same behavior, with $dN/dt$ following a $t^{\alpha-1}$ power-law increase. This observed consistency 
further affirms the validity of our Monte Carlo method. 
Thus, our general observations suggest that the growth curves from effectively contained scenarios \textit{always} ought to be fitted accordingly, by including power-law behavior.                 
This supports the contention that constrained growth curves from global COVID-19 data necessarily require epidemiological analyses that incorporate additional mechanisms, similar to those described in Refs.~\cite{maier,axel}. 

\paragraph*{Summary and conclusions}
In summary, we used a simple two-dimensional random walk Monte Carlo model to study the spread of Covid-19-like infection within a contained population. Apart from proximity based contact, our model has no underlying assumptions about the nature of infection spread or its reproduction number, etc. In addition to establishing similarities with  conventional SIR or SIRS-type models, we show that three growth regimes, corresponding to different levels of containment emerge naturally from our simulations.  
Under stringent conditions, so that only nearest-neighbor connections are allowed, our simulation results show a power-law growth in time, with growth exponents $\alpha = 2.0$-$2.3$, similar to initial COVID-19 data from China~\cite{maier}. 
The determined growth exponents show no apparent dependence on population size or density. Based on available data, this analysis suggests that the containment and mitigation strategies employed/followed in Chinese provinces after the initial outbreak resulted in growth exponents that were close to the smallest limiting value.  
On comparison with data from other countries, we observe that reasonable agreement can be attained by introducing small-world-type connections in the simulation model. We anticipate that such a Monte Carlo approach (and its more generalized versions) will be useful for the evaluation of future strategies in the midst of the present pandemic. 

As concluding remarks, we briefly mention the general similarity between (i) the peripheral growth model~\cite{axel}, (ii) our simulation results for short step-lengths taken by the random walkers, and (iii) the diffusion of particles to distinct sites on a two-dimensional lattice~\cite{larralde} at short time-scales. All these cases show a quadratic growth in time. We further observe that a simple logarithmic correction to our quadratic results (so that $t^2 \to t^2 \ln t$) yields a power-law exponent of about 2.5, in rough agreement with the intermediate values for contained growth, both described here and observed in Refs.~\cite{maier,mingli}. Further investigations into this potential connection present an interesting research problem for both epidemiologists and physicists alike.

\paragraph*{Acknowledgments}We are grateful to Prof.~N.~Barik for fruitful discussions and to Prof. S.\,M.~Bhattacharjee for insightful feedback and directing us to Ref.~\cite{larralde}. 
ST acknowledges funding support from the National Research Foundation, South Africa, under Grant No. 85100.


\bibliography{covid_ST_DPM11}
\end{document}